\documentclass[trackchanges, twocolumn]{aastex701}

\begin{document}

\newcommand{\psr}{PSR~J1922+3745 }
\newcommand{\rcom}{\textcolor{red}}

\title{A Candidate Open Cluster Pulsar: Timing Analysis of PSR J1922+3745 in NGC 6791}

\author[0000-0002-2187-4087]{Xiao-Jin Liu}
\affiliation{Department of Physics, Faculty of Arts and Sciences, Beijing Normal University, Zhuhai 519087, China}
\email{xliu@bnu.edu.cn}

\author[0000-0001-6196-4135]{Ralph P. Eatough}
\affiliation{National Astronomical Observatories, Chinese Academy of Sciences, Beijing 100101, China}
\affiliation{Max-Planck-Institut f\"ur Radioastronomie, Auf dem H\"ugel 69, D-53121 Bonn, Germany}
\email[show]{reatough@nao.cas.cn}

\author[0000-0001-9754-9777]{Zhichen Pan}
\affiliation{National Astronomical Observatories, Chinese Academy of Sciences, Beijing 100101, China}
\affiliation{Guizhou Radio Astronomical Observatory, Guizhou University, Guiyang 550025, China}
\affiliation{Key Laboratory of Radio Astronomy and Technology, Chinese Academy of Sciences, Beijing 100101, China}
\affiliation{College of Astronomy and Space Sciences, University of Chinese Academy of Sciences, Beijing 100049, China}
\email{}

\author[0000-0003-3294-3081]{Matthew Bailes}
\affiliation{Centre for Astrophysics and Supercomputing, Swinburne University of Technology, P.O. Box 218, Hawthorn, VIC 3122, Australia}
\affiliation{OzGrav: The ARC Centre of Excellence for Gravitational Wave Discovery, Hawthorn, VIC 3122, Australia}
\email{mbailes@swin.edu.au}

\author[0000-0002-1056-5895]{Wei-Cong Jing}
\affiliation{National Astronomical Observatories, Chinese Academy of Sciences, Beijing 100101, China}
\email{jingweicong@nao.cas.cn}

\author[0009-0008-0519-084X]{Yong-Sheng Wang}
\affiliation{School of Physics and Astronomy, Beijing Normal University, Beijing 100875, China}
\affiliation{Department of Physics, Faculty of Arts and Sciences, Beijing Normal University, Zhuhai 519087, China}
\email{ying@mail.bnu.edu.cn}

\author[0000-0002-2762-6519]{Xujia Ouyang}
\email{}
\affiliation{College of Physics, Guizhou University,
Guiyang 550025, China}

\author[0000-0002-1086-7922]{Yong Zhang}
\affiliation{School of Physics and Astronomy, Sun Yat-sen University, 2 Daxue Road, Tangjia, Zhuhai, Guangdong Province, China}
\affiliation{Xinjiang Astronomical Observatory, Chinese Academy of Sciences, 150 Science 1-Street, Urumqi, Xinjiang 830011, China}
\affiliation{CSST Science Center for the Guangdong-Hongkong-Macau Greater Bay Area, Sun Yat-Sen University, Guangdong Province, China}
\affiliation{Laboratory for Space Research, The University of Hong Kong, Hong Kong, China}
\email{}

\author[0000-0002-9409-3214]{Rahul Sengar}
\affiliation{Max Planck Institute for Gravitational Physics (Albert Einstein Institute), D-30167 Hannover, Germany}
\affiliation{Leibniz Universität Hannover, D-30167 Hannover, Germany}
\email{rahul.sengar@aei.mpg.de}

\author[0000-0002-5381-6498]{Jianping Yuan}
\affiliation{Xinjiang Astronomical Observatory, Chinese Academy of Sciences, 150 Science 1-Street, Urumqi, Xinjiang 830011, China}
\email{yuanjp@xao.ac.cn}

\author[0000-0002-9786-8548]{Na Wang}
\affiliation{Xinjiang Astronomical Observatory, Chinese Academy of Sciences, 150 Science 1-Street, Urumqi, Xinjiang 830011, China}
\email{}

\author[0000-0001-5105-4058]{Weiwei Zhu}
\affiliation{CAS Key Laboratory of FAST, National Astronomical Observatories, Chinese Academy of Sciences, Beijing 100101, China}
\affiliation{Institute for Frontier in Astronomy and Astrophysics, Beijing Normal University, Beijing 102206, China}
\email{zhuww@nao.cas.cn}

\author[0009-0003-3527-8520]{Peng Jiang}
\email{pjiang@bao.ac.cn}
\affiliation{National Astronomical Observatories, Chinese Academy of Sciences, Beijing 100101, China}
\affiliation{Key Laboratory of Radio Astronomy and Technology, Chinese Academy of Sciences, Beijing 100101, China}

\author[0000-0003-0597-0957]{Lei Qian}
\affiliation{National Astronomical Observatories, Chinese Academy of Sciences, Beijing 100101, China}
\affiliation{Guizhou Radio Astronomical Observatory, Guizhou University, Guiyang 550025, China}
\affiliation{Key Laboratory of Radio Astronomy and Technology, Chinese Academy of Sciences, Beijing 100101, China}
\affiliation{College of Astronomy and Space Sciences, University of Chinese Academy of Sciences, Beijing 100049, China}
\email{}

\author[0009-0000-1929-7121]{Lu Zhou}
\affiliation{School of Physics and Technology, Wuhan University, Wuhan, Hubei 430072, China}
\affiliation{Department of Physics, Faculty of Arts and Sciences, Beijing Normal University, Zhuhai 519087, China}
\email{zhoulugz@163.com}

\author[0000-0003-2516-6288]{He Gao}
\affiliation{School of Physics and Astronomy, Beijing Normal University, Beijing 100875, China}
\affiliation{Institute for Frontier in Astronomy and Astrophysics, Beijing Normal University, Beijing 102206, China}
\email{gaohe@bnu.edu.cn}

\author[0000-0002-3567-6743]{Zong-Hong Zhu}
\affiliation{School of Physics and Astronomy, Beijing Normal University, Beijing 100875, China}
\affiliation{Institute for Frontier in Astronomy and Astrophysics, Beijing Normal University, Beijing 102206, China}
\affiliation{School of Physics and Technology, Wuhan University, Wuhan, Hubei 430072, China}
\email{zhuzh@bnu.edu.cn}

\author[0000-0001-7049-6468]{Xing-Jiang Zhu}
\email[show]{zhuxj@bnu.edu.cn}
\affiliation{Department of Physics, Faculty of Arts and Sciences, Beijing Normal University, Zhuhai 519087, China}
\affiliation{Institute for Frontier in Astronomy and Astrophysics, Beijing Normal University, Beijing 102206, China}

\begin{abstract}
PSR J1922+3745 was recently identified as a radio pulsar toward the old open cluster NGC~6791, raising the prospect of the first pulsar associated with an open cluster. We report FAST follow-up observations that yield a phase-coherent timing solution, a precise position, a measurement of the spin-down rate and the pulsar's polarization properties. PSR J1922+3745 is consistent with an isolated slow pulsar with a characteristic age of 7.8 Myr, comparable to the small population of long-period pulsars found in globular clusters. Motivated by the potential cluster association, we re-process deeper searches of the NGC 6791 field at higher sensitivity but detect no additional pulsars. We also assess whether HI absorption spectroscopy can provide a useful distance constraint and find that such measurements are unlikely to be constraining with currently available sensitivity. Consequently, existing evidence does not yet establish membership in NGC 6791. Further deep searches for additional pulsars with similar dispersion measures in the cluster field will likely be the most direct path to confirming a physical association.

\end{abstract}


\section{Introduction}
Globular clusters (GCs) are exceptionally promising targets for pulsar searches owing to their high stellar densities. To date, around 350 pulsars\footnote{\url{https://www3.mpifr-bonn.mpg.de/staff/pfreire/GCpsr.html}} have been discovered across 46 of the $\sim 170$ known Galactic GCs \citep{VB23}, representing approximately 8\% of the total pulsar population listed in the Australia Telescope National Facility Pulsar Catalogue\footnote{\url{https://www.atnf.csiro.au/research/pulsar/psrcat/}} \citep{MHT+05}. This wealth of discoveries provides a critical resource for investigating neutron star formation histories \citep{IHR+08,VF13}, the dynamical properties of the host clusters \citep{PHI93,PRF+17,FRK+17}, and the potential existence of intermediate-mass black holes residing within them \citep{PRF+17,PSL+17}.

In contrast, open star clusters feature lower stellar densities but outnumber GCs by a factor of more than 20 \citep[e.g.,][]{HR24,LiLu25oc}. Currently, there are no confirmed radio pulsar discoveries within any open cluster, a likely consequence of both suboptimal environments for neutron star formation and retention, as well as a historical lack of dedicated search time. However, because the ages of open clusters can be measured with relatively high accuracy, confirming a pulsar within one would provide a rare and valuable method for determining the true age of the pulsar. Such an independent age calibration would offer vital constraints on pulsar spin-down models \citep[e.g.,][]{LK12,DLC+24} and magnetic field decay processes \citep[e.g.,][]{IPH21}.

Recently, \citet[][hereafter Paper I]{LSB+24} reported the discovery of PSR J1922+3745, a 1.9-second pulsar located in the direction of the old open cluster NGC 6791. The pulsar exhibits a dispersion measure (DM) of 85 pc cm$^{-3}$; when combined with the NE2001 Galactic electron density model \citep{CL02}, this yields a distance estimate consistent with that of NGC 6791, suggesting a potential physical association (Paper I; see also \citealt{FZL+25}). If confirmed, PSR J1922+3745 would be the first radio pulsar found in an open cluster, opening a new chapter for targeted searches that could yield dozens of discoveries across the $\sim$3000 known Galactic open clusters \citep{HR24}. Furthermore, a definitive association would significantly advance our understanding of how slow pulsars can survive in old cluster environments, echoing the puzzles presented by pulsars K and L in the globular cluster M15 \citep{WPQ+24,ZWL+24}.

Despite the compelling implications of a genuine association, the large angular extent of open clusters combined with the dense Galactic pulsar population makes spatial chance associations a significant risk \citep[e.g.,][]{PBP+24,ZYL+25}. Determining whether this pulsar is truly bound to the cluster or merely a line-of-sight coincidence is a critical next step. In this Letter, we present long-term follow-up observations of PSR J1922+3745, yielding the first phase-coherent timing solution, an accurate astrometric position, and a characteristic age to further test the association scenario. We also evaluate the feasibility of placing independent constraints on the pulsar's distance using HI line absorption observations.

The Letter is organized as follows: Section~\ref{sec:observation} describes the follow-up observations. Section~\ref{sec:data_reduction} details the timing analysis and the deeper archival search procedures. In Section~\ref{sec:results}, we discuss the results of our timing solution and HI spectral analysis. Finally, Appendix~\ref{sec:HI_SNR} provides an analytical expression for computing the signal-to-noise ratio of HI line absorption for this source.

\section{Observation} \label{sec:observation}
Due to the weak nature of J1922+3745 ($\sim 7~\mu$Jy at L-band, Paper I), the Five-hundred-meter Aperture Spherical radio
Telescope (FAST, \citealt{NLJ+11,JTH+20}) is one of the few telescopes suitable for follow-up observations.
Using 7.4 hours of FAST Director's Discretionary Time (project ID: DDT2024\_4), we conducted 10 observations between 15 December 2024 and 22 May 2025. 

Table~\ref{tab:observation} presents the basic observation settings. 
In the beginning, we arranged five high-cadence observations in the first week, which facilitated the process of obtaining a timing solution. After that, the observation cadence was reduced to once per month. 
All observations were conducted in tracking mode using the central beam (M01) and data were recorded in search mode, with the exception of the session on 2024--12--16, which used the snapshot mode \citep{HWW+21}.

To obtain a longer timing baseline, legacy data (see lower part of Table~\ref{tab:observation}) from the FAST data archive were incorporated, extending the timing baseline from half a year to more than three years.

\begin{table}
    \centering
    \begin{tabular}{crccc}
    \hline\hline
       Date  &  $t_{\rm int}$ & $t_{\rm samp}$ & $n_{\rm chan}$ & $n_{\rm p}$ \\
       (yy-mm-dd) & (sec) & ($\mu$s) & \\
       \hline       
       2025--05--22  & 1680 & 98.304 & 2048 & 4 \\
       2025--04--22  & 1500 & 49.152 & 4096 & 4 \\
       2025--03--22  & 1500 & 49.152 & 4096 & 4 \\
       2025--02--22  & 1500 & 49.152 & 4096 & 4 \\
       2025--01--22  & 1500 & 49.152 & 4096 & 4 \\
       2024--12--22  & 1500 & 49.152 & 4096 & 4 \\      
       2024--12--20  & 1500 & 49.152 & 4096 & 4 \\
       2024--12--18  & 1500 & 49.152 & 4096 & 4 \\
       2024--12--16  & 4860$^\star$ & 49.152 & 4096 & 4 \\       
       2024--12--15  & 3600 & 49.152 & 4096 & 4 \\
    \hline
       2024--09--12  & 750  & 49.152 & 1024 & 4 \\
       2024--09--01  & 1080 & 49.152 & 2048 & 4 \\
       2022--01--22  & 6900$^\dagger$ & 49.152 & 4096 & 2 \\
    \hline
    \end{tabular}
    \caption{The observation settings of PSR~J1922+3745. The upper part of the table shows the follow-up observations in DDT2024\_4, while the lower part shows the legacy FAST observations, which are included in the timing analysis.
    Note: $^\star$This figure is the total observation length; the tracking time is 1200 sec per pointing, see \cite{HWW+21} for details of snapshot mode. $^\dagger$The figure is the total observation length for OnOff mode; the tracking time is 3435 sec per pointing.}
    \label{tab:observation}
\end{table}

\section{Data reduction}
\label{sec:data_reduction}

\subsection{Timing \& Polarimetry}
To obtain a timing solution, we followed the standard pulsar timing procedures using tools in the \textsc{PSRCHIVE}\footnote{\url{https://psrchive.sourceforge.net/}} software packages \citep{HVM04, sdo+12} unless otherwise stated. 
First, the search mode data were folded into six sub-integrations using the \textsc{dspsr} routine. 
Then, the observation with the highest signal-noise-ratio (S/N) was selected to generate a noise-free template using the \textsc{paas} routine. 
After that, each resulting archive was compared with the template using the \textsc{pat} routine, resulting in four time-of-arrivals (TOAs) per observation.
To minimize the impact of radio frequency interference (RFI), all the archives were cleaned with the \textsc{clfd}\footnote{\url{https://github.com/itachi-gf/clfd/tree/master}} software package \citep{MBC+23} and further examined by eye. 
Finally, the TOAs were analyzed using \textsc{tempo2} \citep{EHM06}. 
Considering the limited length of the timing baseline, we only fit for position (RA and Dec), spin frequency ($f$) and its time derivative ($\dot{f}$), while using the optimized DM from \textsc{pdmp} and keeping the DM constant. 
The obvious outliers were removed during the fitting, and clock files from the FAST observatory\footnote{\url{https://github.com/NAOC-pulsar/FAST_ClockFile}} were also used to minimize the impact of potential clock drifts.
The reference epoch was set to the midpoint of the observational baseline.

The polarimetric analysis also follows the standard procedures using PSRCHIVE. For approximately the first minute of an observation, a noise diode, modulated by a square wave of period $0.201\,{\rm s}$, is fired to inject a signal that can be used to calibrate the instrumental response over the observing band. By applying the calibration solutions we have investigated the intrinsic polarization properties of PSR~J1922+3745. For observations where the calibration signal was not initiated, due to instrumental or observational logistical issues, the closest calibration observation in time was used to calibrate the data. 

\subsection{More in depth pulsar searches}

The discovery of a wider population of pulsars toward NGC~6791, at similar DMs, would strongly indicate their association with the cluster. Using the known DM of PSR~J1922+3745, more sensitive searches of all of the original OnOff survey observations (2021--01--22 in Table~\ref{tab:observation}) have been performed. 
In comparison to Paper~I, the data have been reprocessed with twice as many (4096) frequency channels over the same observational bandwidth of 500\,MHz, the original 8-bit sampling depth, a finer DM step size of 0.046\,pc\, cm$^{-3}$ (about a factor of 2 smaller), analysis of the coherently combined OnOff data\footnote{The full OnOff coherent scan length of 6900\,sec was searched as well as the 3450\,sec On and Off segments. Although the telescope is repointing from the On to Off position during the full scan, for sources that lie in between On Off positions (like PSR~J1922+3745), increased sensitivity is available.} and a larger acceleration range. 
In Paper~I acceleration ranges of $|a|\leq 1\,{\rm m\,s^{-2}}$ were used. 
Here we have employed {\sc accelsearch} in the {\sc PRESTO} \citep{RAN01} software package\footnote{\url{https://github.com/scottransom/presto}} to perform binary pulsar searches. 
A maximum Fourier bin drift $z$ of $|z| = 200$ for all observation lengths analyzed was used. 
For example, an acceleration of $1\,{\rm m\,s^{-2}}$ for a $5\,{\rm ms}$ pulsar corresponds to a $z\sim8$ in a 3450\,sec observation. 
Conversely, a Fourier bin drift of $z=200$ in such an observation would correspond to an acceleration of $\sim 25\,{\rm m\,s^{-2}}$.  
We searched for pulsars across a DM range of $[60, 155]$~pc~cm$^{-3}$ — a span wide enough to encompass the expected DM spread among pulsars physically associated with the cluster, including PSR~J1922+3745. 
This ensures our search is sensitive to multiple cluster members, even if their DMs differ due to spatial distribution and intervening electron density variations.

Searches for single pulses from fast transient sources such as rotating radio transients, giant pulses or fast radio bursts were also performed using \textsc{single\_pulse\_search.py} in \textsc{PRESTO}. 
Box-car filters with widths up to $150\,{t_{\rm samp} \sim7.4\,{\rm ms}}$ were used to record events with intensity $\ge 6\,\sigma$.

In total, 114 separate data segments (On, Off, OnOff) were analyzed. Our data pipeline made use of PulsarX\footnote{\url{https://github.com/ypmen/PulsarX}} \citep{MBCC+23} for efficient radio frequency interference (RFI) mitigation, dedispersion and pulsar candidate folding.

During the manual inspection of folded pulsar candidates, and to easily identify pulsar candidates caused by the harmonics of bright known pulsars - like those of PSR~J1922+3744 itself - we developed a viewing application that automatically computes and displays the period ratio between the candidate and the nearest known pulsar.

\section{Results and Discussion}
\label{sec:results}
\subsection{Timing solution}

\begin{figure}
    \centering
    \includegraphics[trim=25 20 25 20, clip, width=\linewidth]{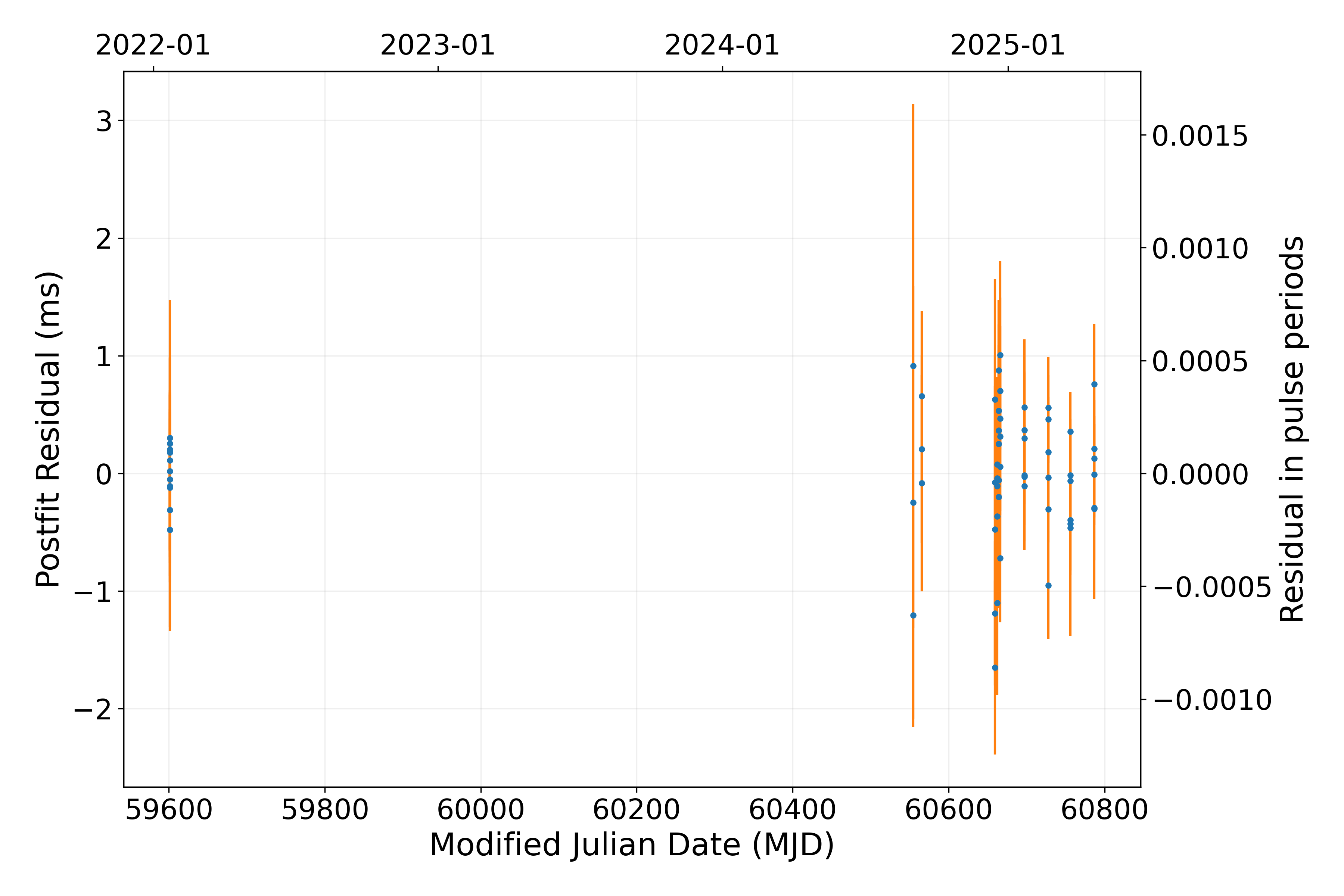}
    \caption{The timing residuals of PSR~J1922+3745. The left $y$-axis is in units of ms, while the right $y$-axis is in units of rotational phase.}
    \label{fig:timing}
\end{figure}

As a result of the timing analysis, a phase-coherent solution was obtained with a post-fit root-mean-square residual of 508~$\mu$s; see Figure~\ref{fig:timing} for the post-fit timing residuals and Table~\ref{tab:parfile} for the post-fit parameters.

Several key points can be drawn from the timing solution.
First, thanks to the 3.25-yr timing baseline, the pulsar position is well constrained with an accuracy of sub-arcsecond.
This is important for giving the correct name to the pulsar, as it is on the boundary between RA=19:21:00 and RA=19:22:00 and can be named either J1921+3745 or J1922+3745, leading to confusion.
Using the most accurate pulsar position, we can confirm the name of the pulsar as J1922+3745.
The accurate position will also facilitate future follow-up observations by providing a higher signal-to-noise ratio (S/N).

Second, PSR~J1922+3745 is likely an isolated pulsar, although we cannot rule out the rare possibility of a wide binary system with an orbital period much longer than the timing baseline (3.25 years), as in the case of PSR~J1024$-$0719 \citep{BJS+16,KKN+16}.

Third, the pulsar spin-down is well constrained to be $\dot{P}=(3.899\pm0.002)\times10^{-15}$~s~s$^{-1}$, giving a characteristic age of $\tau_{\rm c}=7.80$~Myr and a characteristic surface magnetic field strength of $B_{\rm s}=2.77\times10^{12}$~G; see Fig.~\ref{fig:ppdot}.
These parameters thus establish PSR~J1922+3745 as a canonical or slow pulsar.

Note that the estimation above did not correct for other contributions to the apparent $\dot{P}$, although they are all negligible, as shown below. 
First, corrections due to proper motion and Galactic acceleration \citep{SHK70, DT91, LBS18} should not exceed a few times $10^{-18}$~s~s$^{-1}$, if the pulsar is within 9~kpc and moves with a proper motion slower than 200~km/s. 
Second, if the pulsar is associated with the open cluster NGC~6791, then the impact from the cluster is likely more significant \citep{PHI93}, although accurate estimations depend on the angular separation between the cluster center and the pulsar and are proportional to the cluster velocity dispersion.
Following \citet[Eqn. 5]{FRK+17} and using the cluster core radius and angular separation in Paper I, we find an upper limit of $1.5\times10^{-16}$~s~s$^{-1}$ if the velocity dispersion is $<2.5$~mas/yr, which is comparable to the cluster proper motion, thus providing a nice upper limit.
Therefore, the cluster contribution, if they are associated, has negligible impact on both $\tau_{\rm c}$ and $B_{\rm s}$.

Assuming a canonical moment of inertia of $10^{45}$~g~cm$^2$, the corresponding spin-down power is $\dot{E}\sim2.2\times10^{31}$~erg/s, which is less than a thousandth of the empirical threshold of $\sim3\times10^{34}$~erg/s for effective radiation in the gamma-ray band \citep{Tho99,SGC+08}.
The pulsar is thus unlikely to be observable as a gamma-ray source.  

\begin{table}
    \centering
    \footnotesize
    \begin{tabular}{ll}
    \hline
     Right ascension, $\alpha$ (J2000)\dotfill  &  19:22:00.31(2) \\
     Declination, $\delta$ (J2000)\dotfill & +37:45:42.18(8) \\
     Spin frequency, $f$ (Hz)\dotfill & 0.52135 20227 85(2) \\
     Spin period derivative, $\dot{f}~(\times10^{-15}$~Hz$^2$) \dotfill & $-1.0599(6)$ \\
     Reference epoch (MJD)\dotfill & 60193 \\
     Dispersion measure, DM (pc~cm$^{-3}$)\dotfill & 85 \\ 
     \\
     Derived parameters\\
     Spin period, $P$ (s)\dotfill & 1.91808 98055 39(8) \\
     Spin period derivative, $\dot{P}~(\times10^{-15}$~s~s$^{-1}$) \dotfill & $3.899(2)$ \\
     Characteristic age, $\tau_{\rm c}$ (Myr)\dotfill & 7.80 \\
     Surface magnetic field, $B$ ($\times10^{12}$ G)\dotfill & 2.77 \\
     Timing span (yr)\dotfill & 3.25 \\ 
     \hline 
    \end{tabular}
    \caption{Key ephemeris parameters of PSR~J1922+3745 derived from \textsc{tempo2}. Errors quoted in parentheses are 1-$\sigma$.}
    \label{tab:parfile}
\end{table}

\subsection{Polarization properties}
\begin{figure}
    \centering
    \includegraphics[trim=60 40 60 60, clip, width=1.06\linewidth]{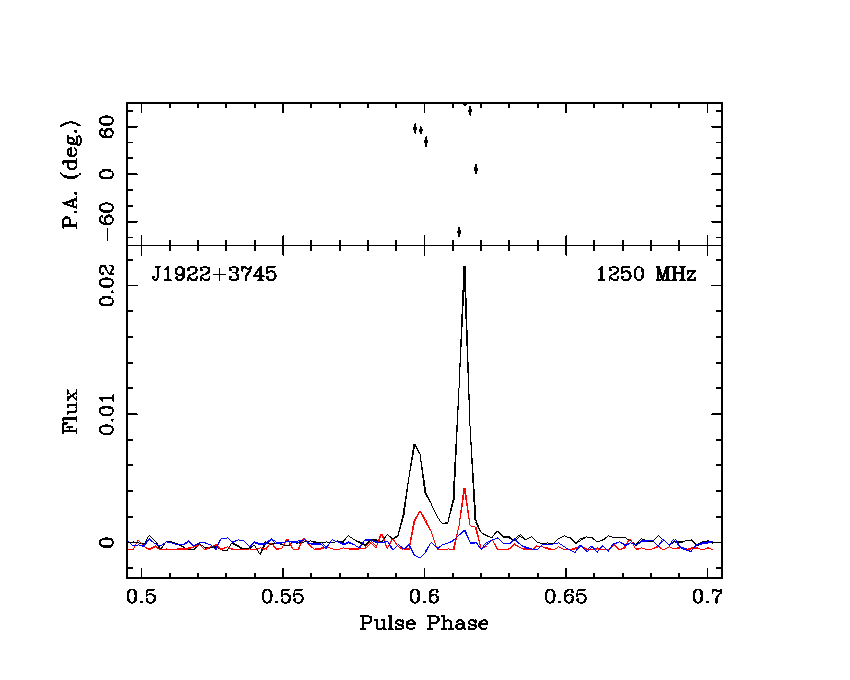}
    \caption{The position angle (top) and profile (bottom) of PSR~J1922+3745, where the black, red and blue curves are for the full power, linear and circular polarizations, respectively.
    For clarity, only the phase region between 0.5 and 0.7 is plotted.}
    \label{fig:profile}
\end{figure}

The final polarization calibrated pulsar profile, formed by combining multiple observations, can be seen in Figure~\ref{fig:profile}. 
After accounting for a Faraday Rotation Measure (RM) of $-23.994\,{\rm rad}\,{\rm m}^{-2}$, PSR~J1922+3745 exhibits a moderate degree of linear polarization ($\sim27\,\%$) and weak circular polarization in both leading and trailing pulse profile components. These levels of polarization are consistent with those seen in the typical pulsar population \citep{LK12}. 
% The upper panel of Figure~1. shows the polarization position angle (PA) swing. 
The polarization properties appear to follow values seen in typical pulsars. 
Other pulsars that are within five degrees of PSR~J1922+3745 show RM values ranging from $\sim -41$ to $103\,{\rm rad}\,{\rm m}^{-2}$ and DMs of $\sim31$ to $103$\,pc\,cm$^{-3}$, showing the RM is broadly consistent with other pulsars in this direction. 
More accurate localization based on both RM and DM will only become possible with the discovery of additional pulsars.

%\subsection{Results of finer search}
\subsection{Results of in depth search}
In total, 53600 candidates were folded and visually inspected, about 470 candidates per pointing-beam.
PSR~J1922+3745 was detected with an S/N of 122, 147 and 188 in the On, Off and OnOff discovery beams respectively, proving the effectiveness of the re-processing pipeline.
In addition, multiple harmonics of the pulsar were easily identified with the help of the auto-computation of period ratios, establishing the viewing application as a handy tool for classifying harmonics, especially in star clusters, which could be crowded with several bright pulsars and dozens of harmonics.
However, no credible new pulsar candidates were identified. 

Multiple single pulses at the DM of $85$~pc~cm$^{-3}$ were also detected throughout the full OnOff scan in the same discovery beam as that of PSR~J1922+3745, but none were detected in other cluster field beams. All detected single pulses are thus from the pulsar, consistent with the results of \cite{FZL+25}. 
A further analysis of the single pulse properties of PSR~J1922+3745 is now underway (Liu et al., in prep).

Using the radiometer equation \citep{LK12} and the default parameters of the FAST 19-beam receiver \citep{JTH+20}, we set a flux limit of 1.2 and 1.7~$\mu$Jy for a duty cycle of 5\% and 10\%, respectively, using the 6900-sec combined OnOff data.

\subsection{HI spectrum}
The observation of HI absorptions provides another way to constrain the distance to a pulsar \citep{FW90,WSX+08,JHH+23,JCL+25}.
The logic behind the method is simple: the intervening HI clouds between us and the pulsar absorb the pulsar emission, thus setting a lower limit on the distance, while the clouds beyond the pulsar do not absorb the pulsar emission. 
Empirically, the absence of absorption lines in the far and bright ($T_{\rm b}>35$~K) HI clouds is used to set an upper limit on the distance \citep{FW90, WSX+08}.
Despite the relatively high Galactic latitude ($b=10$~deg) of PSR~J1922+3745, HI emissions from the direction of the pulsar have been observed in both Parkes \citep{MPC+09} and Effelsberg-Bonn \citep{WKF+16} HI survey\footnote{\url{https://www.astro.uni-bonn.de/hisurvey/AllSky_gauss/index.php}}, with a peak of $\sim 33$~K and an approximate velocity coverage of $[-25, 25]$~km/s.
The low peak temperature thus cannot place any credible upper limits on the pulsar distance.

To set a lower limit on the distance, it is necessary to observe an absorption line in the HI spectrum; however, as shown below, the weak nature of J1922+3745 ($\sim 7~\mu$Jy at L-band, Paper I) makes it unlikely that a reliable HI absorption line can be obtained.
We derived the expression for the signal-to-noise ratio (Equation~\ref{eqn:HI_SNR}) of the HI absorption line of a pulsar in Appendix~\ref{sec:HI_SNR}.
Assuming a line width of 10~km/s (equivalently, $\Delta \nu \sim 50$~kHz), an optical depth of $\tau=0.3$ and an HI temperature of $T_{\rm HI} = 30$~K, and using the intrinsic duty cycle of $\delta=W/P=0.63$\% (Paper I) and the default telescope parameters of FAST \citep{JTH+20}, we find ${\rm S/N}\approx 0.4$ for a tracking length of 6 hr, which is the maximum time allowed by the current observation mode.   

In summary, the observation of the HI spectrum is unlikely to set either a lower or upper limit on the distance to PSR~J1922+3744.
Further considering the difficulty of measuring the parallax and proper motion of the pulsar (Paper I), pulsar searching is the only likely method to establish a definitive association if new pulsars with similar DM are discovered.

\subsection{Comparison with other slow pulsars in globular clusters}

\begin{figure*}
    \centering
    \includegraphics[width=0.7\textwidth]{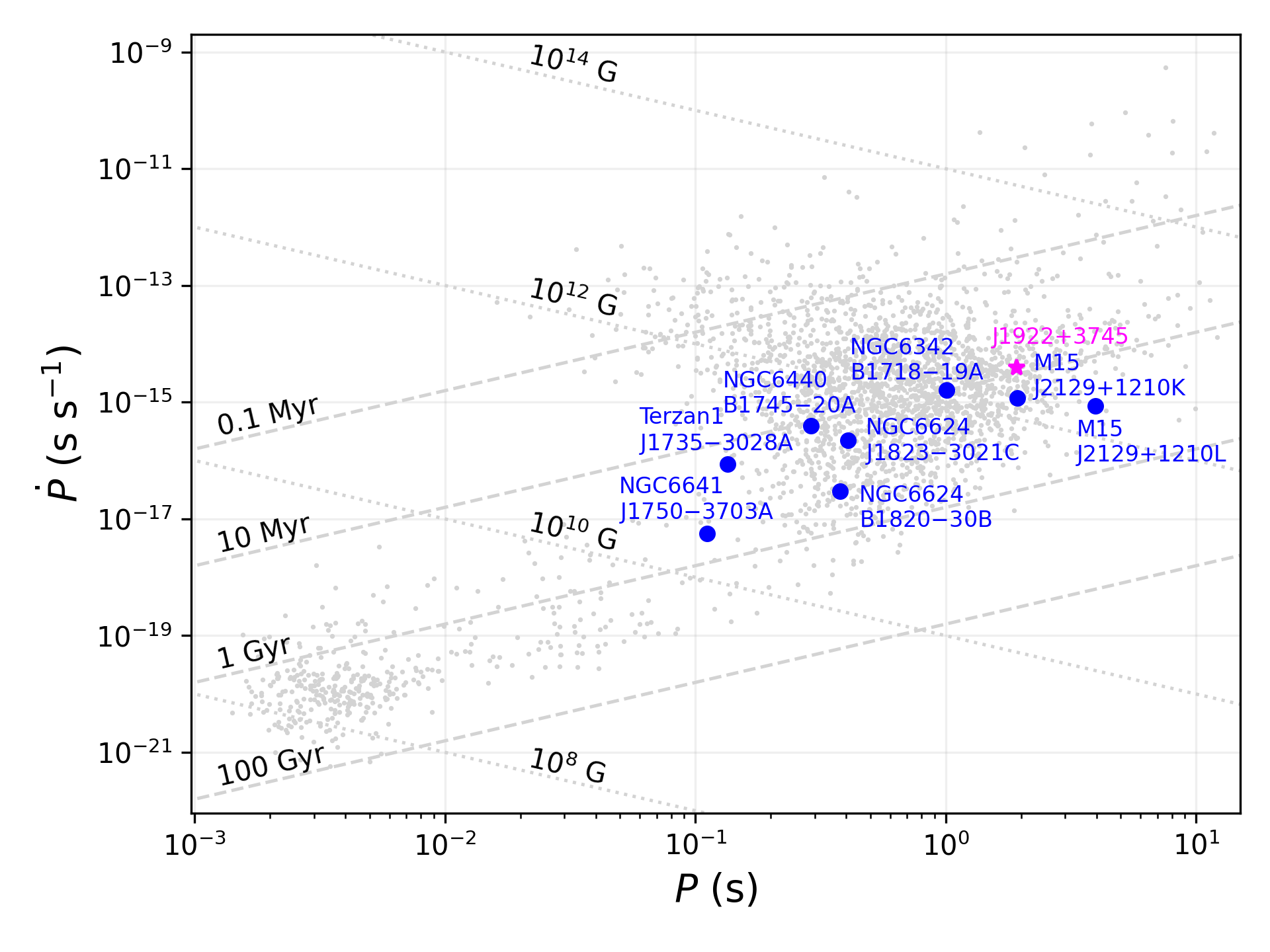}
    \caption{The period-period derivative ($P-\dot{P}$) of PSR~J1922+3745 and the slow pulsars in the globular clusters.
    All other known pulsars are also shown in gray dots for reference.
    The dashed and dotted lines show the characteristic age and surface magnetic field, respectively.
    Note that the error bars of PSR~J1922+3745 and the slow cluster pulsars are too small to be seen.
    \label{fig:ppdot}}
\end{figure*}

A comparison between PSR~J1922+3745 and the slow pulsars in globular clusters may provide useful information regarding the possibility of an association. 
Currently, 13 slow ($P> 100$~ms) pulsars have been discovered in GCs and one additional pulsar, J1823$-$3022, is possibly associated with the GC NGC~6624.
Fig.~\ref{fig:ppdot} shows the $P-\dot{P}$ diagram for 8 of the 13 pulsars, whose apparent $\dot{P}$ are available and greater than $10^{-18}$~s~s$^{-1}$.
To minimize the unknown impact of the cluster potential to $\dot{P}$, we only focus on the slow pulars with a large $\dot{P}$ where the contribution from the cluster potential is minimal or even negligible. 
We see that the spin-down rate of PSR~J1922+3745 is higher than that of all the slow GC pulsars, with the closest pulsar being B1718$-$19A.
Accounting for the difference in spin period, both PSR~J1922+3745 and B1718$-$19A have the similar characteristic age and a comparable surface magntic field.
The $P-\dot{P}$ diagram thus does not contradict with the possible association between PSR~J1922+3745 and NGC~6791.

\section{Conclusions}

We have presented the results of a 3.25-year timing campaign for PSR J1922+3745 using FAST. The resulting phase-coherent timing solution firmly establishes it as an isolated, canonical slow pulsar with a characteristic age of 7.80 Myr and a surface magnetic field of $2.77 \times 10^{12}$ G. The derived spin-down power falls well below the empirical threshold, making observable gamma-ray emission unlikely.

A primary motivation for these follow-up observations was to investigate the potential association between PSR J1922+3745 and the old open cluster NGC 6791. While the pulsar's position is now constrained with sub-arcsecond accuracy, and its placement on the $P-\dot{P}$ diagram is consistent with the small population of slow pulsars found in globular clusters, a definitive physical link remains unconfirmed. To probe this further, we conducted a highly sensitive, targeted pulsar search of the cluster utilizing native-resolution FAST data within a focused dispersion measure range, but this yielded no additional pulsar discoveries.

Finally, we explored the feasibility of establishing independent distance limits for the pulsar via HI absorption spectra. Our analytical assessment indicates that the pulsar's weak L-band flux density ($\sim 7\ \mu\text{Jy}$) and relatively high Galactic latitude make it highly unlikely to secure a reliable HI absorption line detection, even with maximum allowable tracking times. Given the extreme difficulty of obtaining direct parallax and proper motion measurements for this source, we conclude that persistent pulsar searching within NGC 6791 is currently the only viable method to definitively prove an association. The future discovery of additional pulsars with comparable dispersion measures in this direction would provide the ultimate confirmation of this unique open cluster system.

\begin{acknowledgments}
We thank Jinchen Jiang for helpful discussions on the observations of HI.
XJL is supported by the National Science Foundation of China under grant No. 12503046.
X.-J. Zhu is supported by the National Key Research and Development Program of China (No. 2023YFC2206704), the National Natural Science Foundation of China (Grant No.~12203004), the Fundamental Research Funds for the Central Universities, and the Supplemental Funds for Major Scientific Research Projects of Beijing Normal University (Zhuhai) under Project ZHPT2025001.
RPE is supported by the Chinese Academy of Sciences President's International Fellowship Initiative, Grant No. 2021FSM0004.
This work made use of data from FAST (Five-hundred-meter Aperture Spherical radio Telescope; \url{https://cstr.cn/31116.02}. FAST). 
FAST is a Chinese national megascience facility, operated by National Astronomical Observatories, Chinese Academy of Sciences. 
The authors thank the support from the Interdisciplinary Intelligence Super Computer Center of Beijing Normal University at Zhuhai, China.
\end{acknowledgments}

\facilities{FAST}

\software{Astropy \citep{2013A&A...558A..33A, 2018AJ....156..123A,2022ApJ...935..167A}, PSRCHIVE \citep{HVM04}, Tempo2 \citep{EHM06}, CLFD \citep{MBC+23}, PRESTO \citep{RAN01}, PulsarX \citep{MBCC+23}}

\appendix
\section{The S/N of the HI absorption line for a pulsar}
\label{sec:HI_SNR}
The loss of flux density due to the absorption by the HI cloud is 
\begin{equation}
    \Delta S = S_{\rm peak}(1 - \mathrm{e}^{-\tau}),
\end{equation}
where $S_{\rm peak}$ is the peak flux density of the pulsar around 1420~MHz and $\tau$ is the HI optical depth at the line center of absorption. 
Note that the mean flux density, $S_{\rm mean}$, is used more frequently than $S_{\rm peak}$ in the literature of pulsar studies. 
Assuming a top-hat pulse profile, $S_{\rm peak} = S_{\rm mean}P/W=S_{\rm mean}/\delta$, where $P$ and $W$ are the pulse period and width, respectively, and $\delta = W/P$ is usually called duty cycle. 
For clear understanding, we express the flux density in terms of temperature using the gain of a telescope, $G$, i.e. 
\begin{equation}
    \Delta T = G\Delta S = G S_{\rm{mean}} (1 - \mathrm{e}^{-\tau})/\delta. 
\end{equation}

The variance in the absorption baseline is the sum of the pulse-on and pulse-off variance, $\sigma_T^2 = \sigma_{\rm on}^2 + \sigma_{\rm off}^2$. Both $\sigma_{\rm on}$ and $\sigma_{\rm off}$ follow the radiometer equation:
\begin{equation}
    \sigma_{\rm on/off} = \frac{T_{\rm sys}}{\sqrt{n_{\rm p}t_{\rm on/off}\Delta \nu}},
\end{equation}
where $n_{\rm p}$ is the number of independent polarization, $t_{\rm on} + t_{\rm off} = t_{\rm obs}$ is the total observation length, $\Delta \nu$ is the frequency resolution and $T_{\rm sys}$ is the system temperature of HI observation. Therefore, the overall standard deviation, expressed in terms of temperature, is
\begin{equation}
    \sigma_T = \sqrt{\sigma_{\rm on}^2 + \sigma_{\rm off}^2} = \frac{T_{\rm sys}}{\sqrt{n_{\rm p}t_{\rm obs}\Delta \nu}}\sqrt{\frac{1}{\delta(1-\delta)}}.
\end{equation}

Now, the S/N of the HI absorption line for the pulsar is 
\begin{equation}
\label{eqn:HI_SNR}
    {\rm S/N} = \frac{\Delta T}{\sigma_T} = \frac{S_{1420}G\sqrt{n_{\rm p}t_{\rm obs} \Delta\nu}}{T_{\rm rec} + T_{\rm HI}}\sqrt{\frac{P-W}{W}}\Big(1-\mathrm{e}^{-\tau}\Big),
\end{equation}
where we have used $T_{\rm sys} = T_{\rm rec} + T_{\rm HI}$, which is the sum of the receiver and HI temperature \citep{JHH+23}, and renamed 
$S_{\rm mean}$ as $S_{1420}$ to highlight the band of the flux density.

A few points should be noted while using Equation~(\ref{eqn:HI_SNR}) to evaluate the possibility of observing HI absorption lines of pulsars. 
First, $\Delta \nu$ is usually very small (a few kHz) compared with the MHz bandwidth of typical pulsar observations and this is the major reason that limits the application of the method to only a few bright pulsars ($\sim$ a few mJy).
To improve sensitivity, however, one may use the full width of the absorption line. 
That is $\Delta \nu = 1420.4\times\Delta v/c$~MHz, where $c$ is the speed of light in a vacuum and $\Delta v$ is the speed variance of the HI emission line.
For cold HI gas, the line width is around 4 to 10 km/s \citep{DL90, HK07, DSG+09, KK09}, which corresponds to a $\Delta \nu$ of between 19 and 47~kHz.
Second, the intrinsic pulse width, $W_{\rm i}$, is usually smaller than the sampling time, $t_{\rm samp}$, during spectral-line observations. In this case, use $W=(W_{\rm i}^2 + t_{\rm samp}^2)^{1/2}$. 
Third, the HI optical depth is proportional to the bright temperature of the HI clouds. 
For reference, one may assume $\tau\ge 0.3$ when $T_{\rm HI}\ge 35$~K \citep{FW90}. 

\bibliography{J1922}{}
\bibliographystyle{aasjournalv7}

\end{document}